\documentstyle[12pt,a4,sw20jart]{article}
\input tcilatex
\QQQ{Created}{
August 4, 1995
}

\begin{document}

\author{A. Calogeracos \\
%EndAName
NCA Research Consultants, P\ O Box 61147,\\
Maroussi 151 01, Athens, Greece\\
N. Dombey and K. Imagawa\\
Physics and Astronomy Division, University of Sussex,\\
Brighton, BN1 9QH, UK}
\title{SPONTANEOUS FERMION PRODUCTION BY \\
A SUPERCRITICAL POTENTIAL WELL\thanks{%
SUSX.TH/95-48}}
\date{August 4, 1995}
\maketitle

\begin{abstract}
A second quantised theory of electrons and positrons in a deep
time-dependent potential well is discussed. It is shown that positron
production from the well is a natural consequence of Dirac's hole theory
when the strength of the well becomes supercritical. A formalism is
developed whereby the amplitude for emission of a positron of a given
momentum can be calculated. The difference between positron production and
electron-positron pair production is demonstrated. Considerations of the
vacuum charge and of Levinson's theorem are required for a full description
of the problem.

\newpage\ 
\end{abstract}

\section{Introduction}

Gershtein and Zeldovich $\left[ 1\right] $ showed in 1969 that spontaneous
positron production was possible when two heavy bare nuclei with total
charge Z greater than some critical value Z$_c$ collided with each other. Z$%
_c$ is the value of Z for which the lowest 1S state of the hydrogenlike atom
with total charge Z distributed on a nucleus of radius R has energy $E=-m$,
where m is the electron mass. Pomeranchuk and Smorodinskii $\left[ 2\right] $
calculated Z$_c$ = 200 for a nucleus assuming the electrostatic potential is
constant for r$<$R, while Greiner and Pieper $\left[ 3\right] $ calculated Z$%
_c$ = 169 for a homogeneous spherical charge distribution.

Zeldovich and Popov $\left[ 4\right] $ subsequently reviewed the theoretical
basis of this process. (For a recent overview of the problem and of other
related topics the reader may find the monograph $\left[ 5\right] $ useful.)
They were able to set up a self-consistent picture of spontaneous positron
production but from the viewpoint of a first-quantised Dirac theory. They
state, however, that ``at Z $>$ Z$_c$ we have in principle a many-body
problem. This constitutes the third stage of the investigation of the Z $>$ Z%
$_c$ situation, and has not been completed as yet. In an exact formulation,
it is necessary to consider the equations of the electron-positron wave
field with creation and annihilation operators, and to use the second
quantisation formalism''. They go on to say that they predict the results of
experiments on the basis of single particle solutions of the Dirac equation
but that ``such conclusions and predictions require a justification, which
can be obtained only from the exact many-body theory''

This is the problem we have set ourselves in this paper for as far as we
know it has not yet been tackled. For simplicity we have considered
spontaneous fermionic production in the simplest supercritical external
field we could think of, which is a sufficiently deep one-dimensional square
well. We shall find that many of Zeldovich and Popov's results can indeed be
justified. We shall also see that the preparation of the supercritical
potential is an essential feature in the problem and that while it is true
that a static supercritical square well does not radiate $\left[ 6\right] $,
it is also uninteresting because in any experimental situation it is
necessary to create the supercritical potential from sub-critical potentials
(by, for example, colliding two heavy ions together) and thus the
time-dependence of the external field is an essential ingredient.

We first have to elucidate various questions relating to the Dirac equation
in one dimension. In section 2 we consider a Dirac particle in a square
well, write down the spectrum of bound states and give expressions for the
scattering amplitude and the phase shift. We also present the conditions for
criticality. In section 3 we give the second quantized version of the theory
paying particular attention to possible conventions related to the labelling
of states. We review the concept of the vacuum charge that plays a central
role in this discussion. In section 4 we derive Levinson's theorem for the
Dirac equation in one dimension and express the vacuum charge in terms of
phase shifts. In section 5 we illustrate some of the above concepts in the
case of a $\delta $ function potential. In section 6 we apply the methods
and results of the previous sections to the particular case of an external
field which makes the transition from slightly subcritical to slightly
supercritical to slightly subcritical and show in detail that fermionic
radiation can occur. For this it is necessary to arrange for the ground
state of the well to be initially empty, but we demonstrate how we end up
with a filled bound state and a free fermion (which we identify for
convenience with a positron). Second quantization and use of the Bogoliubov
transformation are essential in this treatment.

\section{The One Particle Dirac Equation in One Dimension}

\subsection{The Free Dirac Equation.}

We adopt the convention $\gamma _0=\sigma _3$, $\gamma _1=i\sigma _1$. The
above choice agrees with $\gamma _i\gamma _j+\gamma _j\gamma _i=2g_{ij}.$
The free Dirac Hamiltonian in one dimension is then 
\[
H_0=-\sigma _2p+\sigma _3m 
\]
and the Dirac equation takes the form 
\begin{equation}
(\sigma _1\frac \partial {\partial x}-E\sigma _3+m)\psi =0  \label{frd}
\end{equation}

In what follows ${\bf k}$ stands for the wavevector, $k$ for its magnitude
and $\varepsilon =\left| E\right| =+\sqrt{k^2+m^2}$. We try a plane wave of
the form 
\begin{equation}  \label{frs}
\left( 
\begin{array}{c}
A \\ 
B
\end{array}
\right) e^{ikx-iEt}
\end{equation}

\noindent and substitute in (\ref{frd}). The equation is satisfied by $%
A=ik,B=E-m$ where $E=\pm \varepsilon $. Note that in one dimension the
negative energy solution is obtainable from the positive energy one simply
by replacing $\varepsilon $ by $-\varepsilon $ unlike what happens in three
dimensions. We normalize the 2 dimensional spinor in (\ref{frs}) by
requiring $\int dxu^{\dagger }u=1.$ The positive energy solutions have the
form 
\begin{equation}
N_{+}(\varepsilon )\left( 
\begin{array}{c}
ik \\ 
\varepsilon -m
\end{array}
\right) e^{ikx-i\varepsilon t}  \label{frp}
\end{equation}

\noindent and the negative energy ones 
\begin{equation}
N_{-}(\varepsilon )\left( 
\begin{array}{c}
ik \\ 
-\varepsilon -m
\end{array}
\right) e^{ikx+i\varepsilon t}  \label{frn}
\end{equation}

\noindent where $N_{\pm }(\varepsilon )$ are normalization factors. In what
follows we will either consider the particle in a box of length 2$L$ and
imply periodic boundary conditions at $x=-L$ and $x=L$ or use continuum
normalization$.$ In the former case 
\[
N_{+}(\varepsilon )=\frac 1{\sqrt{2L}\sqrt{2\varepsilon (\varepsilon -m)}}%
\text{ , }N_{-}(\varepsilon )=\frac 1{\sqrt{2L}\sqrt{2\varepsilon
(\varepsilon +m)}} 
\]

\noindent and in the latter 
\[
N_{+}(\varepsilon )=\frac 1{\sqrt{2\pi }\sqrt{2\varepsilon (\varepsilon -m)}}%
\text{ , }N_{-}(\varepsilon )=\frac 1{\sqrt{2\pi }\sqrt{2\varepsilon
(\varepsilon +m)}} 
\]

\noindent Finally we quote the transformation of a wavefunction under
parity: 
\begin{equation}
\psi ^{\prime }(x,t)=\sigma _3\psi (-x,t)  \label{par}
\end{equation}

\subsection{The Dirac Particle in a Square Well:\ Bound States.}

Consider a square well of depth $-V$ extending from $x=-a$ to $x=a.$ The
electron Hamiltonian is 
\[
H_0=-\sigma _2p-V+\sigma _3m 
\]
where $V$ is the zeroth component of $A_\mu $. The electron charge is taken
to be -1. The Dirac equation reads 
\begin{equation}  \label{eqn}
(\sigma _1\frac \partial {\partial x}-(E+V)\sigma _3+m)\psi =0
\end{equation}
Define 
\begin{equation}  \label{p}
p\equiv \sqrt{(E+V)^2-m^2}
\end{equation}
Inside the well try the wavefunction

\[
\left( 
\begin{array}{c}
A\cos px \\ 
B\sin px
\end{array}
\right) 
\]

According to (\ref{par}) this describes a state even under parity.
Substituting in (\ref{frd}) we determine $A,B$ (modulo a factor): 
\begin{equation}  \label{boev}
\psi _{even}=\left( 
\begin{array}{c}
p\cos px \\ 
(E+V-m)\sin px
\end{array}
\right)
\end{equation}

\noindent for $-a\leq x\leq a$. For $x\geq a$ it can easily be checked that 
\begin{equation}
\psi _{even}=s\left( 
\begin{array}{c}
-\kappa \\ 
E-m
\end{array}
\right) e^{-\kappa x}  \label{boev1}
\end{equation}

\noindent with 
\begin{equation}
\kappa =\sqrt{E^2-m^2}  \label{kappa}
\end{equation}
satisfies the Dirac equation. Matching (\ref{boev}) and (\ref{boev1}) at $%
x=a $ and eliminating $s$ we obtain the equation that determines the
spectrum of the even bound states: 
\begin{equation}
\tan ap=\sqrt{\frac{(m-E)(E+V+m)}{(m+E)(E+V-m)}}  \label{sev}
\end{equation}

\noindent Note that it is implicit in (2.7) that for a subcritical potential
(i.e. $V<2m$; see below) permissible bound states (even or odd) satisfy $%
E>m-V$. In a similar way we can determine the wavefunctions of the odd bound
states. It turns out that for $-a\leq x\leq a$%
\begin{equation}
\psi _{odd}=\left( 
\begin{array}{c}
-p\sin px \\ 
(E+V-m)\cos px
\end{array}
\right)  \label{bod}
\end{equation}

\noindent and for $x>a$%
\begin{equation}
\psi _{odd}=s^{\prime }\left( 
\begin{array}{c}
-\kappa \\ 
E-m
\end{array}
\right) e^{-\kappa x}  \label{bod1}
\end{equation}

\noindent Matching (\ref{bod}) and (\ref{bod1}) at $x=a$ and eliminating $%
s^{\prime }$ we obtain the equation that determines the spectrum of the odd
bound states: 
\begin{equation}
\tan ap=-\sqrt{\frac{(m+E)(E+V-m)}{(m-E)(E+V+m)}}
\end{equation}

\noindent These results are well known and can be found, for example, in
reference [3].

\subsection{The Dirac Particle in a Square Well: Scattering States.}

Consider a wave incident from the left. The corresponding wavefunction is 
\begin{equation}  \label{sa}
\left( 
\begin{array}{c}
ik \\ 
E-m
\end{array}
\right) e^{ikx}+B\left( 
\begin{array}{c}
-ik \\ 
E-m
\end{array}
\right) e^{-ikx}
\end{equation}

\noindent for $x<a,$%
\begin{equation}
C\left( 
\begin{array}{c}
ip \\ 
E+V-m
\end{array}
\right) e^{ipx}+D\left( 
\begin{array}{c}
-ip \\ 
E+V-m
\end{array}
\right) e^{-ipx}  \label{sb}
\end{equation}

\noindent for $-a\leq x\leq a$ and 
\begin{equation}
F\left( 
\begin{array}{c}
ik \\ 
E-m
\end{array}
\right) e^{ikx}  \label{sc}
\end{equation}
for $x\geq a$. To calculate the wavefunction we require continuity at $x=-a$
and at $x=a$. The coefficients $B,C,D,F$ are calculated in \cite{ima}. We
only give the results for the phase shift. Define 
\[
F(E)=\left| F(E)\right| e^{i\delta (E)} 
\]

\noindent and 
\begin{equation}
\gamma =\frac kp\frac{E+V+m}{E+m}  \label{gamma}
\end{equation}
Then 
\begin{equation}
\delta (E)=\arctan \left( \frac{1+\gamma ^2}{2\gamma }\tan 2pa\right) -2ka
\label{delta}
\end{equation}

\noindent In what follows the phase shift will sometimes be written with two
arguments, i.e. $\delta (E,V)$, the second referring to the particular value
of the potential. Whenever we write $\delta \left( E,0\right) $ we
understand the limit $\delta \left( E,V\rightarrow 0\right) $. The algebra
leading to (2.19) is right regardless of the sign of the energy. Recalling
that $\epsilon \equiv \left| E\right| $ the phase shifts for positive and
negative energy will be written as $\delta _{+}\left( \epsilon ,V\right)
,\,\delta _{-}\left( \epsilon ,V\right) $ respectively. The phase shift is
defined modulo a multiple of $\pi .$ We choose 
\begin{equation}
\delta (E,0)=0  \label{delo}
\end{equation}
(if $V=0$ then $k\sim p$ , $\gamma \rightarrow 1$ and (\ref{delta})
satisfies $\left( 2.19\right) $)$.$ We also write

\begin{equation}  \label{delinf}
\delta (\pm \infty ,V)=\pm 2Va
\end{equation}

\noindent Indeed if $E\rightarrow +\infty $ (with $V,m$ fixed) then $k\sim
E,p\sim E$, (\ref{gamma}) yields $\gamma \sim 1$ and from (\ref{delta}) $%
\delta \sim 2pa-2ka\rightarrow 2Va.$ Similarly for $E\rightarrow -\infty $ , 
$\gamma \sim 1$ and $\delta \sim 2pa-2ka\rightarrow -2Va.$ Note that $(\ref
{delinf})$ satisfies $\delta (\infty )+\delta (-\infty )=0$ as shown in 3
dimensions by Ma and Ni [8]). We also have 
\[
\delta (\pm \infty )=\mp \int_{-\infty }^\infty V(x)dx 
\]

\noindent which is analagous to the 3-dimensional result \cite{mani} 
\[
\delta (\pm \infty )=\mp \int_0^\infty V(r)dr 
\]

\noindent It must be emphasized that this result is in contrast to what
happens in nonrelativistic quantum mechanics where the phase shift at high
energy always tends to zero (modulo a multiple of $\pi ).$ The absence of
multiples of $\pi $ in the right hand side of (2.20), (2.21) is not a matter
of convention. That would be the case had we specified e.g. $\delta \left(
+\infty ,\,V\right) $ only. Having however fixed the phase shift at one end
of the energy range the rest should follow. It is better to regard (2.20),
(2.21) as dictated by physical considerations as well. For $V\rightarrow 0$
we expect the results of the Dirac equation to coincide with those of the
Schroedinger equation and (2.20), (2.21) satisfy this requirement (without
any multiples of $\pi $). Notice that the negative energy phase shift makes
sense in the context of nonrelativistic quantum mechanics as well: it
corresponds to the scattering of a positron of momentum $-{\bf k}.$

We now wish to make a statement concerning the threshold values $\delta (\pm
m,V).$ For a certain fixed $V$ consider positive ( negative) energy states
very near threshold, i.e. $E\rightarrow m+$ ($E\rightarrow -m-$). Then the
quantity $\gamma $ defined in (\ref{gamma}) has small absolute value and is
positive (negative). Hence from (\ref{delta}) we conclude that 
\begin{equation}
\delta (m,V)=\frac \pi 2+n(V)\pi  \label{delmp}
\end{equation}

\begin{equation}  \label{delmn}
\delta (-m,V)=-\frac \pi 2+n^{\prime }(V)\pi
\end{equation}

\noindent where $n\left( V\right) ,\,n^{\prime }\left( V\right) $ are
integers. Notice that at this stage no statement is made about $n\left(
V\right) ,\,n^{\prime }\left( V\right) $. In particular the values at
threshold $n\left( 0\right) ,\,n^{\prime }\left( 0\right) $ are unknown; see
section 4 for details. Although the above equations have the look of
Levinson's theorem in nonrelativistic quantum mechanics (see e.g. \cite
{schiff}, p.354) in the case of the Dirac equation it is the sum $%
n+n^{\prime }$ that is connected to the number of bound states [8]. We
consider this point in more detail in section 4. However the occurrence of $%
\dfrac \pi 2$ in (\ref{delmp}) and (\ref{delmn}) has the same origin as the $%
\dfrac \pi 2$ in the corresponding equation in \cite{schiff}, namely the
presence of a half-bound state in one-dimension [9].

We are particularly interested in the case where there is no reflection and
so the transmission coefficient is a maximum. These we call, following Bohm$%
\,\left[ 10\right] $, transmission resonances. Setting $B=0$ and matching
the above expressions at the boundaries we get four equations and
eliminating $C,D,F$ we obtain the equation determining the spectrum of the
resonances 
\begin{equation}
2pa=N\pi  \label{res}
\end{equation}

\noindent This relation connecting wavelength and dimension of the well is
familiar from the Fabry-Perot etalon. The similarity in the mathematical
treatment of Fabry-Perot and a Dirac square well was first pointed out in
reference [6]. We should emphasize, however, that transmission resonances
are not resonances in the normal sense; the phase shift does not increase
through $\pi /2$ as it does with a proper resonance.

A further point concerning the relation between bound states and
transmission resonances may be in order. Consider a transmission resonance
of energy $E$ with $k$ the incident wavevector and $p$ the wavevector
between the walls. Using the boundary conditions one can easily show that 
\begin{equation}
\frac CD=e^{i2p^{\prime }a}  \label{cdres}
\end{equation}

\noindent The above equation together with (\ref{res}) yield $C=\pm D.$
Substituting in (\ref{sb}) we get that the transmission resonance
wavefunction between the walls for $N=1$ has the form (\ref{boev}). In other
words if we consider an almost critical potential then the wavefunction of
the lowest lying ({\sl even) }bound state and the wavefunction between the
walls of the upper ($N=1)$ resonance approach each other as the potential
varies and the energies $E_{bound}$ and $E_{res}$ approach $-m$ from above
and below respectively. Hence the overlap integral betwen the two
wavefunctions is sizeable.

We define positive energy spinor wavefunctions $u_{(+)}({\bf k},x)$ given by
(\ref{sa}), (\ref{sb}), (\ref{sc}) multiplied by $N_{+}(\varepsilon )$ and
negative energy ones $u_{(-)}({\bf k},x)$ replacing $E$ by $-E$ multiplied
by $N_{-}(\varepsilon ).$ Note that $u_{(+)}^{\dagger
}u_{(-)}=u_{(-)}^{\dagger }u_{(+)}=0$ and 
\begin{equation}
\int dxu_{(+)}^{\dagger }({\bf k},x)u_{(+)}({\bf k}^{\prime },x)=\int
dxu_{(-)}^{\dagger }({\bf k},x)u_{(-)}({\bf k}^{\prime },x)=\delta ({\bf k}-%
{\bf k}^{\prime })  \label{orth}
\end{equation}
$u^{\dagger }({\bf k},x)=u(-{\bf k},x)$ for both kinds of spinors. These
wavefunctions refer to scattering states. We also have to include bound
states. The spinor wavefunction corresponding to the $j$th bound state is
denoted by $u_j(x)$ and can be read off (\ref{boev}), (\ref{boev1}), (\ref
{bod}), (\ref{bod1}) (at present the notation does not distinguish between
even and odd bound states). $u_j$ is assumed to be normalized $\int
dxu_j^{\dagger }u_j=1$ and is orthogonal to the continuum wavefunctions.

Since parity is a good quantum number it is convenient to introduce spinor
wavefunctions that transform under parity in a definite way. Using (\ref{par}%
) define parity even and odd spinors corresponding to positive energies: 
\begin{equation}  \label{parp}
u_{(+)e}(k,x)\equiv \frac 1{\sqrt{2}}\left( u_{(+)}({\bf k},x)-\sigma
_3u_{(+)}(-{\bf k},x)\right)
\end{equation}
\begin{equation}  \label{parn}
u_{(+)o}(k,x)\equiv \frac 1{\sqrt{2}}\left( u_{(+)}({\bf k},x)+\sigma
_3u_{(+)}(-{\bf k},x)\right)
\end{equation}

\noindent Replacing the subscript (+)\ by (-) we obtain the corresponding
expressions for negative energy spinors.

\subsection{Criticality}

Let us now address the notion of criticality that plays a crucial role in
the discussion. It is quite clear that for a small value of the potential
equation (\ref{sev}) always has a solution corresponding to an even bound
state. Suppose that the potential deepens gradually. At some stage the
energy of the bound state crosses zero; the implications of this effect on
the vacuum charge will be discussed later. For greater values of the
potential the energy approaches $-m$ and for a critical value $V_1^c$ (the
subscript 1 refers to the fact that this is the first bound state that
disappears) the bound state merges with the negative energy continuum
states. Whether other bound states have appeared in the meantime is
considered presently. The wavevector $p$ between the walls is related to the
potential by (\ref{p}) and solving for $V$ we get 
\begin{equation}
V=\sqrt{p^2+m^2}-E  \label{v}
\end{equation}

\noindent It is clear from (\ref{sev}) that when 
\begin{equation}
ap=\frac \pi 2  \label{p1cr}
\end{equation}

\noindent the even bound state is at $-m$ and subsequently disappears. This
corresponds to a critical value of the potential 
\begin{equation}
V_1^c=\sqrt{\frac{\pi ^2}{4a^2}+m^2}+m  \label{v1cr}
\end{equation}

\noindent Comparing (\ref{p1cr}) with (\ref{res}) we see that the bound
state wavefunction goes over to the $N=1$ transmission resonance
wavefunction. On the other hand it is clear from (2.14) that when $ap=\dfrac
\pi 2$ a new{\sl \ odd} bound state appears at $E=m$. This corresponds to
the value of the potential

\begin{equation}
V_{odd1}=\sqrt{\frac{\pi ^2}{4a^2}+m^2}-m  \label{v1odd}
\end{equation}

\noindent When $pa=\pi $ the second {\sl even }bound state appears at 
\begin{equation}
V_{even,2}=\sqrt{\frac{\pi ^2}{a^2}+m^2}-m  \label{v2even}
\end{equation}

Again for $pa=\pi $ the first {\sl odd }bound state disappears corresponding
to the potential 
\begin{equation}  \label{v2cr}
V_2^c=\sqrt{\frac{\pi ^2}{a^2}+m^2}+m
\end{equation}

\noindent and to the $N=2$ transmission resonance. It should by now be clear
how the pattern repeats itself.

\newpage\ 

\section{Second Quantization.}

Let $N_{+},N_{-}$ be the number of positive and negative bound states
respectively. We expand $\psi $ in terms of the continuous spectrum
wavefunctions (\ref{parp}), (\ref{parn}) and of the bound states:

\begin{equation}  \label{sec}
\begin{array}{c}
\psi
(x,t)=\sum_k\{a_e(k,0)u_{(+)e}(k,x)e^{-iEt}+a_o(k,0)u_{(+)o}(k,x)e^{-iEt}+
\\ 
\\ 
+c_e^{\dagger }(k,0)u_{(-)e}(k,x)e^{iEt}+c_o^{\dagger
}(k,0)u_{(-)o}(k,x)e^{iEt}\}+ \\ 
\\ 
+\{\sum_{j=1}^{N_{+}}b_j(0)u_j(x)e^{-iE_jt}+\sum_{j=1}^{N_{-}}d_j^{\dagger
}(0)u_j(x)e^{-iE_jt}\}
\end{array}
\end{equation}

\smallskip\ 

\noindent Operators $a^{\dagger },a$ create and annihilate travelling
electrons; $c^{\dagger },c$ are the corresponding ones for positrons.
Operators $b_j$ $(b_j^{\dagger })$ annihilate (create) bound electrons
whereas $d_j$ ($d_j^{\dagger }$) annihilate (create) bound positrons. The
use of the ($\dagger $) in (\ref{sec}) is dictated by the sign of the
exponential and conforms to current literature. The Hermitean conjugate
expansion is 
\begin{equation}
\begin{array}{c}
\psi ^{\dagger }(x,t)=\sum_k\{a_e^{\dagger }(k,0)u_{(+)e}^{\dagger
}(k,x)e^{iEt}+a_o^{\dagger }(k,0)u_{(+)o}^{\dagger }(k,x)e^{iEt}+ \\ 
\\ 
+c_e(k,0)u_{(-)e}^{\dagger }(k,x)e^{-iEt}+c_o(k,0)u_{(-)o}^{\dagger
}(k,x)e^{-iEt}\}+ \\ 
\\ 
+\{\sum_{j=1}^{N_{+}}b_j^{\dagger }(0)u_j^{\dagger
}(x)e^{iE_jt}+\sum_{j=1}^{N_{-}}d_j(0)u_j(x)e^{iE_jt}\}
\end{array}
\label{secc}
\end{equation}

\smallskip\ 

\noindent where we took into account the reality of $u_j.$ The standard
anticommutation relations are obeyed 
\begin{equation}
\begin{array}{c}
\{a_{e,o}(k,t),a_{e,o}^{\dagger }(k^{\prime },t\}=\delta _{kk^{\prime
}}\delta _{eo}\text{ , }\{b_i(t),b_j^{\dagger }(t)\}=\delta _{ij}\text{ } \\ 
\\ 
\{c_{e,o}(k,t),c_{e,o}^{\dagger }(k^{\prime },t\}=\delta _{kk^{\prime
}}\delta _{eo}\text{ , }\{d_i(t),d_j^{\dagger }(t)\}=\delta _{ij}\text{ }
\end{array}
\label{anti}
\end{equation}

\noindent We work in the Heisenberg picture throughout: The time dependence
is carried by operators whereas state vectors are time independent. However
basis ket vectors (and in particular the vacuum) are time dependent (see
e.g. \cite{sak}). The vacuum $\left| 0\right\rangle $ is defined by 
\begin{equation}
a_{e,o}(k)\left| 0\right\rangle =c_{e,o}(k)\left| 0\right\rangle =b_i\left|
0\right\rangle =d_i\left| 0\right\rangle =0  \label{vg1}
\end{equation}

\noindent It must be stressed however that the definition of the vacuum is
to some extent a matter of convention. A complete statement on the physics
of the problem is made when one defines both the vacuum and also the state
vector of the system. We will elaborate this point later on.

The total charge is defined by (N.B. according to our conventions the
electron charge is $-$1)\ 
\begin{equation}  \label{ch1}
Q(t)\equiv \int dx\rho (x,t)=-\frac 12\int dx\left[ \psi ^{\dagger
}(x,t),\psi (x,t)\right]
\end{equation}

\noindent Substituting (\ref{sec}), (\ref{secc}) and using (\ref{anti}) we
get 
\begin{equation}
Q=Q_{norm}+Q_0  \label{ch}
\end{equation}

\noindent where

\begin{equation}  \label{chn}
\begin{array}{c}
Q_{norm}\equiv \sum_k\{-a_e^{\dagger }(k,t)a_e(k,t)+c_e^{\dagger
}(k,t)c_e(k,t)_{}-a_o^{\dagger }(k,t)a_o(k,t)+ \\ 
\\ 
+c_o^{\dagger }(k,t)c_o(k,t)-\sum_{j=1}^{N_{+}}b_j^{\dagger
}b_j+\sum_{j=1}^{N_{-}}d_j^{\dagger }d_j\}
\end{array}
\end{equation}

\noindent and 
\begin{equation}
Q_0\equiv \frac 12\left\{ \sum_k(\text{states with }E>0)-\sum_k(\text{states
with }E<0)\right\}  \label{qo}
\end{equation}

\noindent Expressions (\ref{chn}), (\ref{qo}) are rather formal. To be
precise we have to calculate phase shifts and density of states and
transform the sums to integrals. This is done in section 4 and illustrated
in section 5. Given the definition (\ref{vg1}) of the vacuum we immediately
get 
\begin{equation}
\left\langle 0\right| Q\left| 0\right\rangle =Q_0  \label{vgch}
\end{equation}

\noindent i.e. the vacuum charge turns out to be the spectral asymmetry of
the Hamiltonian. Note that we would not have obtained the connection between 
$Q_0$ and the vacuum asymmetry had we not identified bound states with $E<0$
as positrons or not used the commutator in (\ref{ch1}). Note also that $Q_0$
clearly depends on the potential. The same applies to $Q_{norm}$ although
the notation does not indicate this. The wavefunctions that appear in (3.1),
(3.2) depend on the strength of the potential. By implication the same is
true for the creation and annihilation operators and thus for $Q_{norm}$.

To see precisely how the above formalism should be used let us consider the
electron field interacting with a time dependent potential $V(x,t)$. The
Hamiltonian is given by 
\begin{equation}
{\cal H}={\cal H}(freeDirac)-\frac 12\int dxV(x,t)\left[ \psi ^{\dagger
}(x,t),\psi (x,t)\right]  \label{ham}
\end{equation}

\noindent It is quite clear that (\ref{ham}) conserves $Q$. In order to
avoid extraneous complications let us assume for the moment that the
potential is relatively weak and that there is only one bound state at
energy $E_b$ slightly above zero , i.e. $N_{+}=1,N_{-}=0$. Denote by $%
Q_{0,init}$ the initial vacuum charge. In the present case (\ref{sec}) reads 
\begin{equation}
\begin{array}{c}
\psi
(x,t)=\sum_k\{a_e(k,0)u_{(+)e}(k,x)e^{-iEt}+a_o(k,0)u_{(+)o}(k,x)e^{-iEt}+
\\ 
\\ 
+c_e^{\dagger }(k,0)u_{(-)e}(k,x)e^{iEt}+c_o^{\dagger
}(k,0)u_{(-)o}(k,x)e^{iEt}\}+b(0)u(x)e^{-iE_bt}
\end{array}
\label{sub1}
\end{equation}

\noindent Before going any further we have to specify the state $\left|
S\right\rangle $ of the system. We take all electron and positron scattering
states to be empty and the bound state is also unfilled (in other words this
state coincides with the vacuum state as defined in (\ref{vg1}). Then 
\begin{equation}
a_{e,o}(k)\left| S\right\rangle =c_{e,o}(k)\left| S\right\rangle =0
\label{s}
\end{equation}

\begin{equation}  \label{ss}
b\left| S\right\rangle =0
\end{equation}

\noindent According to (\ref{s}) $\left\langle S\right| Q_{norm}\left|
S\right\rangle =0$ and so 
\begin{equation}
\left\langle S\right| Q\left| S\right\rangle =Q_{0,init}  \label{qinit}
\end{equation}

\noindent We now slowly increase the potential strength until eventually the
bound state energy is found below$E=0$. We denote everything pertaining to
this later time by a ($\symbol{94}$). According to (\ref{sec}) the field now
evolves as 
\begin{equation}
\begin{array}{c}
\hat \psi (x,t)=\sum_k\{\hat a_e(k,0)\hat u_{(+)e}(k,x)e^{-i\hat Et}+\hat
a_o(k,0)\hat u_{(+)o}(k,x)e^{-i\hat Et}+ \\ 
\\ 
+\hat c_e^{\dagger }(k,0)\hat u_{(-)e}(k,x)e^{i\hat Et}+\hat c_o^{\dagger
}(k,0)\hat u_{(-)o}(k,x)e^{i\hat Et}\}+\hat d^{\dagger }(0)\hat u(x)e^{i\hat
E_bt}
\end{array}
\label{sub2}
\end{equation}

\noindent Since the change in potential strength is assumed to be
sufficiently slow and small and since there was no electron occupying the
bound state initiallly, it is clear that there will still be no electron
occupying the bound state.Nevertheless, by the conventional interpretation
of Dirac's theory, the absence of an electron in a state with $E<0$
signifies the presence of a positron which in this case is a bound positron.
For a slow enough change in the potential strength it is also clear that no
radiation is emitted. Thus in terms of the ($\symbol{94})$ operators the
state $\left| S\right\rangle $ satisfies 
\[
\hat a_{e,o}(k)\left| S\right\rangle =\hat c_{e,o}(k)\left| S\right\rangle
=0 
\]

\begin{equation}  \label{dds}
\hat d^{\dagger }\hat d\left| S\right\rangle =\left| S\right\rangle
\end{equation}

\noindent Despite the appearance of a bound positron total charge is
conserved, i.e. $\left\langle S\right| Q\left| S\right\rangle $ is still
equal to $Q_{0,init}$ when $Q$ is expressed in terms of the ($\symbol{94})$
operators. To see this note that the $\hat d^{\dagger }\hat d$ term in $%
Q_{norm}$ (\ref{chn}) now contributes a term +1 (the positron charge)
because of (\ref{dds}). This however is cancelled by a term $-$1 originating
from $Q_0$ simply by counting states; since one bound state has now dived
below zero the vacuum asymmetry is equal to $Q_{0,init}-1$.

Finally it may be of interest to examine the expectation value of the total
charge when, as the potential changes, the energy of the bound state exactly
reaches zero. We can then use either expansion 
\begin{equation}  \label{sub1a}
\begin{array}{c}
\psi
(x,t)=\sum_k\{a_e(k,0)u_{(+)e}(k,x)e^{-iEt}+a_o(k,0)u_{(+)o}(k,x)e^{-iEt}+
\\ 
\\ 
+c_e^{\dagger }(k,0)u_{(-)e}(k,x)e^{iEt}+c_o^{\dagger
}(k,0)u_{(-)o}(k,x)e^{iEt}\}+bu(x)
\end{array}
\end{equation}

\noindent or 
\begin{equation}
\begin{array}{c}
\psi
(x,t)=\sum_k\{a_e(k,0)u_{(+)e}(k,x)e^{-iEt}+a_o(k,0)u_{(+)o}(k,x)e^{-iEt}+
\\ 
\\ 
+c_e^{\dagger }(k,0)u_{(-)e}(k,x)e^{iEt}+c_o^{\dagger
}(k,0)u_{(-)o}(k,x)e^{iEt}\}+d^{\dagger }u(x)
\end{array}
\label{sub1b}
\end{equation}

\noindent The use of $b$ or $d^{\dagger }$ in (\ref{sub1a}) and (\ref{sub1b}%
) respectively is purely a matter of taste since in the absence of an energy
exponential our convention dictates nothing. Assuming again that the bound
state level has an electron vacancy (\ref{sub1a}) entails 
\begin{equation}
b\left| S\right\rangle =0  \label{zero1}
\end{equation}

\noindent whereas (\ref{sub1b}) entails 
\begin{equation}
d^{\dagger }d\left| S\right\rangle =\left| S\right\rangle  \label{zero2}
\end{equation}

\noindent In both cases scattering states are vacant 
\[
a_{e,o}(k)\left| S\right\rangle =c_{e,o}(k)\left| S\right\rangle =0 
\]

\noindent Notice however that the two conventions have different
implications for the value of $Q_0$ when the level is at exactly zero
energy. According to (\ref{zero1}) the zero mode is counted as an electron
state hence $Q_0$ of (\ref{qo}) gets a $+\dfrac 12$ contribution due to
normal ordering of $b$ and $b^{\dagger }$. In contrast according to (\ref
{zero2}) the zero mode is counted as a positron state hence $Q_0$ gets a $%
-\dfrac 12$ contribution due to normal ordering of $d$ and $d^{\dagger }$.
However physical results do not change. When we calculate $\left\langle
S\right| Q_{norm}\left| S\right\rangle $ then according to the first
convention there is no contribution from the zero mode because of (\ref
{zero1}). On the other hand according to the second convention there is a +1
contribution because of (\ref{zero2}) and the presence of the $d^{\dagger }d$
term in $Q_{norm}$. Thus the total contribution to $\left\langle S\right|
Q\left| S\right\rangle $ due to the zero mode is $+\dfrac 12$ as before.
Notice that when the bound state is slightly above zero then its
contribution to $Q_0$ is $+\dfrac 12$ (according to (\ref{qo}) and its
contribution to $\left\langle S\right| Q_{norm}\left| S\right\rangle $ is
zero according to (\ref{ss}). Thus the contribution of the bound state to $%
\left\langle S\right| Q\left| S\right\rangle $ depends smoothly on the
potential.

\newpage\ 

\section{Phase Shifts at Threshold and Levinson's Theorem and the Vacuum
Charge.}

We now discuss Levinson's theorem $\left[ 13\right] $, $\left[ 14\right] ,$ $%
\left[ 15\right] $, $\left[ 16\right] $, for a one-dimensional system since
it is intimately related to the discussion of the vacuum charge on which our
interpretation of positron radiation is based. We follow the strategy used
by Barton $\left[ 12\right] $ in his discussion of Levinson's theorem for
the Schroedinger equation in one dimension and enclose the system in a box
of length 2$L$ with periodic boundary conditions 
\begin{equation}
\psi \left( -L\right) =\psi \left( L\right)
\end{equation}

\noindent (Note that if one compares what follows with reference $12$ one
should interchange the roles of integers $n$ and $v$.) We consider even and
odd states separately. An odd state of positive energy has asymptotically
the form 
\begin{equation}
\psi _{odd}\left( x\rightarrow \pm \infty \right) =\left( 
\begin{array}{c}
ik\sin \left( kx\pm \Delta _{o+}\right) \\ 
\left( \epsilon -m\right) \cos \left( kx\pm \Delta _{o+}\right)
\end{array}
\right)
\end{equation}

\noindent and an even one 
\begin{equation}
\psi _{even}\left( x\rightarrow \pm \infty \right) =\left( 
\begin{array}{c}
ik\cos \left( kx\pm \Delta _{e+}\right) \\ 
\left( \epsilon -m\right) \sin \left( kx\pm \Delta _{e+}\right)
\end{array}
\right)
\end{equation}

\noindent where we neglect normalization factors (the subscript $\pm $
denotes the energy sign). Negative energy states are obtained by replacing $%
\epsilon $ by $-\epsilon $ as already mentioned in section 1. The phase
shifts introduced in section 2 are connected to the above by 
\begin{equation}
\delta _{+}\left( \epsilon \right) =\Delta _{e+}\left( \epsilon \right)
+\Delta _{o+}\left( \epsilon \right)
\end{equation}

\begin{equation}
\delta _{-}\left( \epsilon \right) =\Delta _{e-}\left( \epsilon \right)
+\Delta _{o-}\left( \epsilon \right)
\end{equation}

\noindent Condition (4.1) implies

\begin{equation}
k\left( \nu _{e,\pm }\right) L+\Delta _{e\pm }\left( k_\nu \right) =\nu
_{e,\pm }\pi
\end{equation}

\begin{equation}
k\left( \nu _{o,\pm }\right) L+\Delta _{o\pm }\left( k_\nu \right) =\nu
_{o,\pm }\pi
\end{equation}

\noindent $\nu $ being integer. It is clear from (4.2), (4.3) that the
wavevectors $k$ are strictly positive; this will impose restrictions on $\nu 
$. In the continuum limit calculations similar to the ones outlined in
section 2 yield 
\begin{equation}
\Delta _e\left( \pm m\right) =\frac \pi 2+n_{e,\pm }\left( 0\right) \pi
\end{equation}

\begin{equation}
\Delta _o\left( \pm m\right) =n_{o,\pm }\left( 0\right) \pi
\end{equation}

\noindent (the argument in the above relations indictes that $V\rightarrow 0$%
). In the free case relations (4.6), (4.7) are still valid with no $\Delta $
appearing. For V small enough, the system reduces to that of the
Schroedinger equation discussed in $[12]$. In that case Barton shows that
for counting purposes it is sufficient to consider the number of positive
energy states for a free particle to be integers $\nu _{e,+}$, $\nu _{o,+}$
ranging from a minimum value 
\begin{equation}
\nu _{\min ,free}=1
\end{equation}

\noindent to some large cut-off value $N$. We repeat the process for
negative energy states.We thus have 2$N$ even solutions and 2$N$ odd ones
(counting both positive and negative energy states). Suppose that we now
switch on the potential keeping it arbitrarily small. We know that the
number of positive energy even scattering states decreases by 1 since there
is one even bound state appearing. We have the same number of odd scattering
states and the same number of negative energy scattering states. The total
number of states is conserved. In other words 
\begin{equation}
\nu _{\min ,o,-}=\nu _{\min ,o,+}=\nu _{\min ,e,-}=1
\end{equation}

\begin{equation}
\nu _{\min ,e,+}=2
\end{equation}

\noindent It should also be observed that for any value of $\nu _{\min }$
the wavevector $k\left( \nu _{\min }\right) $ goes to zero in the large $L$
limit. We are thus entitled to replace $\Delta _{e,o}\left( k\left( \nu
_{\min ,\pm }\right) \right) $ by the appropriate threshold value $\Delta
\left( \pm m\right) $. Substituting (4.8) in (4.6) we get 
\begin{equation}
k\left( \nu _{e,+}\right) L=-\frac \pi 2-n_{e,+}(0)\pi +\nu _{e,+}\pi
\end{equation}

\begin{equation}
k\left( \nu _{e,-}\right) L=-\frac \pi 2-n_{e,-}\left( 0\right) \pi +\nu
_{e,-}\pi
\end{equation}

\noindent Similarly 
\begin{equation}
k\left( \nu _{o,+}\right) L=-n_{o,+}\left( 0\right) \pi +v_{o,+}\pi
\end{equation}

\begin{equation}
k\left( \nu _{o,-}\right) L=-n_{o,-}\left( 0\right) \pi +\nu _{o,-}\pi
\end{equation}

\noindent From (4.11), (4.12) and from the fact that $k$ is strictly
positive we deduce that 
\begin{equation}
n_{e,+}\left( 0\right) =1
\end{equation}

\begin{equation}
n_{e,-}\left( 0\right) =0
\end{equation}

\noindent In other words 
\[
\Delta _e\left( m\right) =\frac{3\pi }2\,,\Delta _e\left( -m\right) =\frac
\pi 2 
\]

\noindent Via a similar argument 
\[
n_{o,+}\left( 0\right) =n_{o,-}\left( 0\right) =0 
\]

\noindent and 
\begin{equation}
\Delta _o\left( m\right) =\Delta _o\left( -m\right) =0
\end{equation}

\noindent This concludes the question of the determination of phase shifts
at threshold mentioned after (4.5). Note that $\Delta _e\left( \pm m\right) $
has a discontinuity of $\dfrac \pi 2$ at $V=0$ just as it does for the one
dimensional Schroedinger equation since an attractive one dimensional
potential no matter how weak always has at least one bound state.

We can now turn on the potential gradually: then for specific values of $V$
(cf. the discussion at the end of section 2) a scattering state
(alternatively even or odd) crosses $m$ and becomes bound. Since one
scattering state is lost it is clear that $\nu _{\min ,e,+}$ or $\nu _{\min
,o,+}$ respectively increases by 1. From the requirement that $k$ be
positive we deduce from (4.11) and (4.17) that $n_{e,+}\left( V\right) $ and
n$_{o,+}\left( V\right) $ denote the number of even and odd states
respectively that have crossed $m$ for the given value of the potential
(including the even bound state that exists just below $m$ for an
arbitrarily small value of the potential). Similarly as the potential
deepens further, bound states cross $-m$ and join the continuum. The fact
that more negative energy scattering states become available means that $\nu
_{\min ,e,-}$ and $\nu _{\min ,o,-}$ decrease and from (4.18), (4.16) we
deduce that $n_{e,-}$ and $n_{o,-}$ decrease. Since (cf. (4.16), (4.18))
they start at zero for vanishing potential we conclude that $n_{e,-}$ and $%
n_{o,-}$ are negative and that the absolute values $\left| n_{e,-}\left(
V\right) \right| $ and $\left| n_{o,-}\left( V\right) \right| $ represent
the number of even and odd (ex) bound states that have crossed $-m$ for the
particular value of $V$. Simple bookkeeping then yields that $n_{e,+}\left(
V\right) +n_{e,-}\left( V\right) $ and $n_{o,+}\left( V\right)
+n_{o,-}\left( V\right) $ denote the number of even and odd bound states
respectively. From (4.8) and (4.9) we get 
\begin{equation}
n_{bound,even}=\frac 1\pi \left( \Delta _e\left( m,V\right) +\Delta _e\left(
-m,V\right) \right) -1
\end{equation}

\begin{equation}
n_{bound,odd}=\frac 1\pi \left( \Delta _o\left( m,V\right) +\Delta _o\left(
-m,V\right) \right)
\end{equation}

\noindent The above relations constitute Levinson's theorem in the present
problem.

The connection between the number of bound states and the jumps of phase
shifts at threshold by $\pi $ can be seen directly in the case of the square
well examined in section 2. Return to expression (2.19) for the phase shift,
set $k\sim 0,E\sim m$ so that the quantity $\dfrac 1\gamma $ in (2.19) tends
to infinity regardless of the strength of the potential and vary $V$. For
some particular value $V_{new}$ a new bound state crosses $E=m$. Recall that
according to the analysis of section 2.4 tan $2pa$ as a function of $V$
vanishes at $V_{new}$. From (2.7) it is also clear that tan $2pa$ is an
increasing function of $V$, so since $V$ is increasing, tan $2pa$ is crosses
zero from $below$. Thus according to (2.19) 
\[
\delta \left( m,\,V_{new}-0\right) =\arctan \left( -\infty \right) 
\]

\[
\delta \left( m,\,V_{new}+0\right) =\arctan \left( \infty \right) 
\]

\noindent hence when a new bound state appears $\delta \left( m\right) $
jumps by $\pi $, i.e. $n$ in (2.22) increases by 1. In other words $n$
counts the number of bound states that appear (including the bound state
that exists just below $m$ for an arbitrarily small $V$). By a similar
reasoning when a bound state crosses zero and disappears (for {\it increasing%
} $V$), $\delta \left( -m\right) $ jumps by $-\pi $, i.e. $n^{\prime }$
decreases by 1. Hence for an attractive potential $n^{\prime }$ is negative
and its absolute value counts the number of bound states that disappeared.

Before we move to the determination of the vacuum charge let us point out
the constraints that (4.7), (4.9) impose on the minimum value of the
wavevector $k$ (for $L$ large but finite). Setting $\Delta =0$ in the above
relations we get that for a free field in all cases (positive or negative
energy, even or odd parity) 
\begin{equation}
k_{\min ,free}=\frac \pi L
\end{equation}

\noindent corresponding to $\nu =1$. For a very weak potential (4.7), (4.9)
in conjunction with (4.15), (4.16), (4.19) yield (again in all cases) 
\begin{equation}
k_{\min }=\frac \pi {2L}
\end{equation}

\noindent To obtain the expression for the vacuum charge we convert sums
over states using the standard expression that follows from (4.2), (4.3)

\begin{equation}
\sum (states)=\frac 1\pi \int^\infty dk\left( L+\frac{d\delta }{dk}\right)
\label{sum}
\end{equation}

\noindent where the lower limit will be stated presently. Separating the
contributions of odd and even states (3.8) gives 
\begin{eqnarray}
Q_0 &=&\frac 12\{\frac 1\pi \int_{k_{\min }}^\infty dk\left( L+\frac{d\Delta
_{e+}(E)}{dk}\right) +n_{bound,even,+} \\
&&+\frac 1\pi \int_{k_{\min }}^\infty dk\left( L+\frac{d\delta _{o+}(-E)}{dk}%
\right) +n_{bound,odd,+}-\frac 1\pi \int_{k_{\min }}^\infty dk\left( L+\frac{%
d\Delta _{e-}(E)}{dk}\right) -  \nonumber \\
&&-n_{bound,even,-}-\frac 1\pi \int_{k_{\min }}^\infty dk\left( L+\frac{%
d\Delta _{o-}(e)}{dk}\right) -n_{bound,odd,-}  \nonumber
\end{eqnarray}

\noindent with $k_{\min }$ given by (4.23). It should be noticed that
although $k_{\min \text{ }}$tends to zero in the $L\rightarrow \infty $
limit this is counterbalanced in the integral by the factor $L$. Although
this is irrelevant for our purposes it is in general essential if we want to
ensure conservation of the total number of states. The fact that the lower
limits of integration in (4.25) are all identical allows the $L$
proportional terms to cancel out after replacing $\infty $ by a cut-off
wavevector $K$ (this is true even after a more careful regularization. The $%
dk$ integration over the $\Delta $ functions is performed trivially and
brings in the functions $\delta _{+}(\epsilon )$, $\delta _{-}(\epsilon )$
via (4.4), (4.5). For convenience we define the number of bound states
according to the energy sign and regardless of parity 
\[
N_{+}\equiv n_{bound,even,+}+n_{bound,odd,+} 
\]

\[
N_{-}\equiv n_{bound,even,-}+n_{bound,odd,-} 
\]

\noindent Then Levinson's theorem can be trivially rewritten 
\[
N_{+}+N_{-}=\frac 1\pi \left( \delta _{+}(m)+\delta _{-}(m)\right) -1 
\]

\noindent We finally get 
\begin{equation}
Q_0=\frac 12\left\{ \frac 1\pi \left( \delta _{+}(\infty ,V)-\delta
_{+}\left( m,V\right) -\delta _{-}(\infty ,V)+\delta _{-}(m,V)\right)
+N_{+}-N_{-}\right\}  \label{qoph}
\end{equation}

\noindent In the particular case of a square well (2.21), (4.26) yield 
\begin{equation}
Q_0=\frac 12\left\{ \frac{4Va}\pi +\frac 1\pi \left( -\delta _{+}\left(
m,V\right) +\delta _{-}\left( m,V\right) \right) +N_{+}-N_{-}\right\}
\end{equation}

It may be appropriate at this point to enlarge on the role played by $%
k_{\min }$ introducted in (4.22), (4.23) and explain the meaning of the
statement following (4.25). Suppose that the potential is extremely weak and
that we have one (just) bound state. It is then natural to assume that the
total number of positive energy states (scattering plus bound) is the same
as before. Equation (4.24) gives 
\begin{eqnarray}
\sum \left( \text{{\it positive states}, }V\rightarrow 0\right) &=&\frac
L\pi \int_{k_{\min }}^Kdk+\frac 1\pi \int_{k_{\min }}^Kdk\frac{d\delta }{dk}%
+1= \\
&&\frac{Lk}\pi -\frac L\pi k_{\min }+\frac 1\pi \delta \left( k_{\min
}\right) +1  \nonumber
\end{eqnarray}

\noindent For a sufficiently weak potential $\delta \left( K\right) $ can be
neglected. As already explained for large $L\delta \left( k_{\min }\right)
\sim \delta \left( 0\right) =\dfrac{3\pi }2$. If we were to neglect the role
of $k_{\min }$ (i.e. set $k_{\min }=0$) we would get 
\begin{equation}
\sum (\text{{\it positive states, }}V\rightarrow 0)=\frac{LK}\pi -\frac 12
\end{equation}

\noindent Given the first term in (4.29) would be there even in the free
case it is clear that the above result and the idea of dropping $k_{\min }$
are absurd: (4.29) implies that switching on the potential creates {\it half}
a state. However if we take (4.23) into account we get 
\begin{equation}
\sum (\text{{\it positive states, }}V\rightarrow 0)=\frac{LK}\pi -\frac L\pi
.\,\,\frac \pi {2L}-\frac 1\pi .\frac{3\pi }2+1=\frac{LK}\pi -1
\end{equation}

\noindent On the other hand for {\it strictly} {\it zero} potential strength 
$k_{\min }$ is given by (4.22), hence 
\[
\sum (\text{{\it positive states, }}V=0)=\frac{LK}\pi -\frac L\pi .\,\,\frac
\pi L-\frac \pi L=\frac{LK}\pi -1 
\]

\noindent in manifest agreement with (4.30).

Let us focus on the behaviour of $Q_0$ for small changes of the potential.
When the potential changes slightly and a new bound state appears the phase
shifts $\delta (\pm \infty )=\pm 2Va$ change smoothly$.$ The jump of $\delta
(m)$ by $\pi $ in (\ref{qoph}) is counterbalanced by the increase in $N_{+}$
by 1. Thus $Q_0$ also changes smoothly. On the other hand if for a slight
change of the potential a bound state crosses zero then the phase shifts
behave smoothly, $N_{+}$ decreases by 1 and $N_{-}$ increases by 1. Hence $%
Q_0$ decreases abruptly by 1. This however should not necessarily be
construed as a physical discontinuity. Recall for example the argument after
(\ref{dds}): For the particular state $\left| S\right\rangle $ considered
there the decrease in $Q_0$ is counterbalanced by a corresponding increase
in $\left\langle S\right| Q_{norm}\left| S\right\rangle $ since the empty
bound electron state now appears as an occupied positron state with charge
+1. Thus the total charge has no discontinuity.

Let us see this last point in somewhat greater detail. Suppose that one
discrete level (the lowest lying even state in the present case) crosses
zero for $V=V_0.$ (For definiteness assume that $a$ is such that the first
odd state has already appeared; it can be checked that this is indeed
feasible for say $a=.7.)$ Then at $V=V_0-$ the even bound state is just
above zero and we assume it to be empty. The vacuum charge is obtained from
(4.27) with $N_{+}=2,$ $N_{-}=0,$ $n=1,$ $n^{\prime }=0$%
\begin{equation}
Q_0=\frac{2V_0a}\pi  \label{qolo}
\end{equation}

\noindent and the expectation value of the total charge is 
\[
\left\langle S\right| Q(V_0-)\left| S\right\rangle =\left\langle S\right|
Q_{norm}(V_0-)\left| S\right\rangle +Q_0 
\]

\noindent where the argument $\left( V_0-\right) $ serves to remind that
creation and annihilation operators depend on $V$. Consider a very small
change in the potential so that $V=V_0+$. Since the even bound state has
dived below zero it is now a positron state and according to the hole theory
(and to our conventions) the absence of an electron amounts to the presence
of a positron. So in terms of operators pertaining to $V=V_0+$ the state $%
\left| S\right\rangle $ defined by (\ref{s}) satisfies 
\begin{equation}
d^{\dagger }(V_0+)d(V_0+)\left| S\right\rangle =\left| S\right\rangle
\label{do}
\end{equation}

Since the potential changed infinitesimally no radiation has been emitted.
Hence according to (\ref{chn}) 
\[
\left\langle S\right| Q_{norm}(V_0+)\left| S\right\rangle =\left\langle
S\right| Q_{norm}(V_0-)\left| S\right\rangle +1 
\]

\noindent where the last term is due to the presence of $d_{even}^{\dagger
}d_{even}$ in $Q_{norm}$ and simply reflects the fact that the positron
carries charge +1. The important point is that charge is conserved. Indeed
now $N_{+}=1,N_{-}=1$ and according to (\ref{qoph2}) $Q_0$ has decreased by
1: 
\begin{equation}
Q_0\text{=}\frac{2V_0a}\pi -1  \label{qocrit}
\end{equation}
The above reflects the connection between vacuum charge and spectral
asymmetry. The charge expectation value $\left\langle S\right| Q\left|
S\right\rangle $ rests unchanged during the crossing.

\newpage\ 

\section{An Example: The $\delta $ function Potential.}

We consider the attractive (for electrons) potential $-\lambda \delta (x)$.
We define this potential as a square well in the limit $V\rightarrow \infty
,a\rightarrow 0$ so that $Va$ is kept finite and $\lambda =2Va.$ (This
definition avoids the problems of definition discussed in reference [14].
The advantage of the $\delta $ function limit lies in the fact that we can
obtain simpler closed formulae for some of the quantities treated previously.

We first turn to the determination of the scattering amplitude and the phase
shift. For a wave incident from the left 
\begin{equation}  \label{soll}
\psi (x<0)=\left( 
\begin{array}{c}
ik \\ 
E-m
\end{array}
\right) e^{ikx}+R\left( 
\begin{array}{c}
-ik \\ 
E-m
\end{array}
\right) e^{-ikx}
\end{equation}

\begin{equation}  \label{solr}
\psi (x>0)=F\left( 
\begin{array}{c}
ik \\ 
E-m
\end{array}
\right) e^{ikx}
\end{equation}

A solution to (\ref{eqn}) is clearly furnished by 
\begin{equation}  \label{solexp}
\psi (x)=\int_{x_0}^xdx\exp (i\sigma _2(E+\lambda \delta (x)-m))\psi (x_0)
\end{equation}

\noindent for some arbitrary $x_0$. Applying the above equation for $%
x_0=0-,x=0+$ we connect (\ref{soll}), (\ref{solr}) 
\begin{equation}
\psi (0+)=e^{i\lambda \sigma _2}\psi (0-)  \label{con}
\end{equation}

or 
\[
\left( 
\begin{array}{c}
ik \\ 
E-m
\end{array}
\right) +R\left( 
\begin{array}{c}
-ik \\ 
E-m
\end{array}
\right) =(\cos \lambda +i\sin \lambda \sigma _2)F\left( 
\begin{array}{c}
ik \\ 
E-m
\end{array}
\right) 
\]

We thus get two equations for $F,R$. After trivial algebra 
\[
F(E)=\frac k{k\cos \lambda -iE\sin \lambda } 
\]

This leads to 
\begin{equation}  \label{deld}
\delta (E)=\arctan \left( \frac Ek\tan \lambda \right)
\end{equation}

Let us compare the above result to (\ref{delta}). To this end the $%
V\rightarrow \infty $ limit in section 2 should be taken from the start.
Then clearly $2pa=\lambda .$ Again in the same limit and after some algebra
we get 
\[
\frac{1+\gamma ^2}{2\gamma }\rightarrow \frac Ek 
\]

Thus (\ref{delta}) and (\ref{deld}) agree. In the high energy limit clearly 
\begin{equation}  \label{delinfd}
\delta (\pm \infty )=\pm \lambda
\end{equation}

\noindent (note that this agrees with the convention (\ref{delo})). It also
agrees with result (\ref{delinf}). At threshold we get 
\begin{equation}
\delta (m)=\frac \pi 2+n(\lambda )\pi  \label{delmpd}
\end{equation}

\begin{equation}  \label{delmnd}
\delta (-m)=-\frac \pi 2+n^{\prime }(\lambda )\pi
\end{equation}

\noindent The analysis of the previous section still holds and entails $%
n(0)=n^{\prime }(0)=0$ at threshold$.$

We now turn to the spectrum of the bound states. The wavefunctions are of
the form 
\[
\left( 
\begin{array}{c}
-\kappa \\ 
E-m
\end{array}
\right) e^{-\kappa \left| x\right| } 
\]

\noindent modulo a sign depending on parity. Applying (\ref{con}) we get for
even bound states 
\begin{equation}
\left( 
\begin{array}{c}
-\kappa \\ 
E-m
\end{array}
\right) =\left( 
\begin{array}{cc}
\cos \lambda & \sin \lambda \\ 
-\sin \lambda & \cos \lambda
\end{array}
\right) \left( 
\begin{array}{c}
-\kappa \\ 
E-m
\end{array}
\right)  \label{boevd}
\end{equation}

\noindent and for odd bound states 
\begin{equation}
\left( 
\begin{array}{c}
-\kappa \\ 
E-m
\end{array}
\right) =-\left( 
\begin{array}{cc}
\cos \lambda & \sin \lambda \\ 
-\sin \lambda & \cos \lambda
\end{array}
\right) \left( 
\begin{array}{c}
-\kappa \\ 
E-m
\end{array}
\right)  \label{boodd}
\end{equation}

\noindent Using relation (\ref{kappa}) between $\kappa $ and $E$ we get the
equations determining the spectrum. For even bound states 
\begin{equation}
\tan \frac \lambda 2=\sqrt{\frac{m-E}{m+E}}  \label{sevd}
\end{equation}

\noindent and for odd bound states 
\begin{equation}
\tan \frac \lambda 2=-\sqrt{\frac{m+E}{m-E}}  \label{sodd}
\end{equation}

\noindent Equations (\ref{sevd}), (\ref{sodd}) are identical to (\ref{sev}),
(\ref{sod}) in the $V\rightarrow \infty $ limit. Both give the equation for
the spectrum 
\begin{equation}
E=m\cos \lambda \text{sign}(\sin \lambda )  \label{spec}
\end{equation}
Thus even bound states appear when $\dfrac \lambda 2=N\pi $ and disappear
when $\dfrac \lambda 2=\dfrac \pi 2+N\pi $. Odd bound states appear when $%
\dfrac \lambda 2=-\dfrac \pi 2+N\pi $ and disappear when $\dfrac \lambda
2=N\pi .$ A distinctive feature of the $\delta $ function potential is that
whenever a bound state of a certain parity appears a state of the opposite
parity disappears (and vice versa). For example when $\lambda =\pi $ the
original (even) bound disappears and the first odd one appears, at $\lambda
=2\pi $ the odd state disappears and the second even state appears etc.. In
other words $\lambda {\it \limfunc{mod}}\pi $ counts the number of bound
states (excluding the first one that exists for arbitrarily small $\lambda $%
) that have crossed $E=m.$ These results may also be obtained from the
analysis of section 2.4 in the limit $a\rightarrow 0$ (thus essentially
dropping $m$ everywhere). Finally from (\ref{spec}) we deduce that a
discrete level crosses zero when $\lambda =\dfrac \pi 2+N\pi .$

Let us examine the vacuum charge in the $\delta $ function case. Suppose
that $N$ bound states (on top of the original half bound state existing for
arbitrarily small $\lambda $) have crossed $m$, i.e. $\lambda =N\pi +\lambda
^{\prime },$ 0$\leq \lambda ^{\prime }<\pi $. Then $\delta (\infty )=\lambda 
$ , $\delta (-\infty )=-\lambda ,$ $\delta (m)=\dfrac{3\pi }2+N\pi ,$ $%
\delta (-m)=-\dfrac \pi 2-N\pi .$ If the bound state is above zero, i.e. $%
N_{+}=1$ , $N_{-}=0$ (\ref{qoph}) gives 
\begin{equation}
Q_0=\frac{\lambda ^{\prime }}\pi  \label{qod1}
\end{equation}

\noindent In the same way if $N_{+}=0$ , $N_{-}=1$%
\begin{equation}
Q_0=\frac{\lambda ^{\prime }}\pi -1  \label{qod2}
\end{equation}

\noindent Thus at $\lambda =\dfrac \pi 2$ (when the crossing takes place)
the vacuum carge jumps from $+\dfrac 12$ to $-\dfrac 12$ , the difference
being correctly equal to 1. As explained towards the end of section 3.1 the
value of the vacuum charge {\sl at }$\lambda =\dfrac \pi 2$ depends on our
convention concerning the labelling of the zero mode.

\newpage\ 

\section{The Transition from a Subcritical to a Supercritical Potential and
Positron Emission.}

We now finally consider the transition from a subcritical to a supercritical
potential. We assume that the potential starts at a slightly subcritical
value $V_{sub}$, makes an abrupt change to a supercritical value $V_{\sup
er} $ at time $t_1$, stays fixed at this value until time $t_2$ and then
makes an abrupt jump back to $V_{sub}.$ By assumption there is a positron
occupying the bound state for if we had started with an initially vacant
electron bound state (for a weak potential) then as the potential became
stronger the bound state would have crossed $E=0$ and then the absence of an
electron in that state is now interpreted as the presence of a positron. The
implicit assumption is that the initial electron vacancy (or equivalently
positron prescence) persists during the switching on process from $V_{weak}$
to $V_{sub}$. Wavefunctions pertaining to the supercritical potential and
corresponding creation and annihilation operators will be denoted by a ($%
\sim $), the time argument of the latter ranging from $t_1$ to $t_2.$

The crucial observation is that although the potential may change abruptly
at $t=t_1$ the field $\psi (x,t)$ is continuous. Thus we can expand $\psi
(x,t_1)$ in terms of operators pertaining both to $t=t_1-$ and to $t=t_1+.$
For $t=t_1-$ the expansions read 
\begin{equation}  \label{psim}
\begin{array}{c}
\psi (x,t_1)=\sum_k\{a_e(k,t_1)u_{(+)e}(k,x)+a_o(k,t_1)u_{(+)o}(k,x)+ \\ 
\\ 
+c_e^{\dagger }(k,t_1)u_{(-)e}(k,x)+c_o^{\dagger
}(k,t_1)u_{(-)o}(k,x)\}+d_{even}^{\dagger
}(t_1)u_{even}(x)+b_{odd}(t_1)u_{odd}(t_1)
\end{array}
\end{equation}

\begin{equation}  \label{psimd}
\begin{array}{c}
\psi ^{\dagger }(x,t_1)=\sum_k\{a_e^{\dagger }(k,t_1)u_{(+)e}^{\dagger
}(k,x)+a_o^{\dagger }(k,t_1)u_{(+)o}^{\dagger }(k,x)+ \\ 
\\ 
+c_e(k,t_1)u_{(-)e}^{\dagger }(k,x)+c_o(k,t_1)u_{(-)o}^{\dagger
}(k,x)\}+d_{even}(t_1)u_{even}(x)+b_{odd}^{\dagger }(t_1)u_{odd}(t_1)
\end{array}
\end{equation}

\smallskip\ 

\noindent and for $t=t_1+$ 
\begin{equation}
\begin{array}{c}
\psi (x,t_1)=\sum_k\{\tilde a_e(k,t_1)\tilde u_{(+)e}(k,x)+\tilde
a_o(k,t_1)\tilde u_{(+)o}(k,x)+ \\ 
\\ 
+\tilde c_e^{\dagger }(k,t_1)\tilde u_{(-)e}(k,x)+\tilde c_o^{\dagger
}(k,t_1)\tilde u_{(-)o}(k,x)\}+\tilde b_{odd}(t_1)\tilde u_{odd}(t_1) \\ 
\end{array}
\label{psf}
\end{equation}

\smallskip\ 

\begin{equation}  \label{psfd}
\begin{array}{c}
\psi ^{\dagger }(x,t_1)=\sum_k\{\tilde a_e^{\dagger }(k,t_1)\tilde
u_{(+)e}^{\dagger }(k,x)+\tilde a_o^{\dagger }(k,t_1)\tilde
u_{(+)o}^{\dagger }(k,x)+ \\ 
\\ 
+\tilde c_e(k,t_1)\tilde u_{(-)e}^{\dagger }(k,x)+\tilde c_o(k,t_1)\tilde
u_{(-)o}^{\dagger }(k,x)\}+\tilde b_{odd}^{\dagger }(t_1)\tilde
u_{odd}^{\dagger }(t_1)
\end{array}
\end{equation}

\smallskip\ 

Expansion (\ref{psf}) can be inverted to give 
\begin{equation}  \label{a}
\tilde a_e(k,t_1)=\int dx\tilde u_{(+)e}^{\dagger }(k,x)\psi (x,t_1)
\end{equation}

\begin{equation}  \label{b}
\tilde c_e^{\dagger }(k,t_1)=\int dx\tilde u_{(-)e}^{\dagger }(k,x)\psi
(x,t_1)
\end{equation}

Similar expressions hold for the odd operators and for the bound state ones.
Expressions for the Hermitean conjugates can be obtained by taking the
complex conjugates of the right hand sides of (\ref{a},\ref{b}). We invoke
continuity and substitute (\ref{psf}) in (\ref{a}), (\ref{b}). We notice
that we get overlap integrals of the form $\int dx\tilde u_{(\pm
)e,o}^{\dagger }(k,x)u_{(\pm )e,o}(k^{\prime },x)$. Given that outside the
well the wavefuctions are plane waves such integrals are proportional to $%
\delta _{kk^{\prime }}\delta _{eo}$. We thus get

\begin{equation}  \label{aai}
\tilde a_e(k,t_1)=A_ka_e(k,t_1)+B_kc_e^{\dagger
}(k,t_1)+F_kd_{even}^{\dagger }(t_1)
\end{equation}

\begin{equation}  \label{bbi}
\tilde c_e^{\dagger }(k,t_1)=G_ka_e(k,t_1)+L_kc_e^{\dagger
}(k,t_1)+M_kd_{even}^{\dagger }(t_1)
\end{equation}

\noindent where 
\begin{equation}
\begin{array}{c}
\begin{array}{c}
A_k=\int dx\tilde u_{(+)e}^{\dagger }(k,x)u_{(+)e}(k,x),B_k=\int dx\tilde
u_{(+)e}^{\dagger }(k,x)u_{(-)e}(k,x) \\ 
\\ 
F_k=\int dx\tilde u_{(+)e}^{\dagger }(k,x)u_{even}(x),\text{ }G_k=\int
dx\tilde u_{(-)e}^{\dagger }(k,x)u_{(+)e}(k,x)
\end{array}
\\ 
\\ 
L_k=\int dx\tilde u_{(-)e}^{\dagger }(k,x)u_{(-)e}(k,x),\text{ }M_k=\int
dx\tilde u_{(-)e}^{\dagger }(k,x)u_{even}(x)
\end{array}
\label{coef}
\end{equation}

\noindent Expressions for $\tilde a_e^{\dagger }(k,t_1)$ and $\tilde
c_e(k,t_1)$ are obtained by taking the Hermitean conjugates of (\ref{aai}), (%
\ref{bbi}) respectively. The corresponding equations for the odd modes of
course involve the odd rather than the even bound state because of parity 
\begin{equation}
\tilde a_o(k,t_1)=A_k^{\prime }a_o(k,t_1)+B_k^{\prime }c_o^{\dagger
}(k,t_1)+F_k^{\prime }b_{odd}(t_1)  \label{aaii}
\end{equation}

\begin{equation}  \label{bbii}
\tilde c_o^{\dagger }(k,t_1)=G_k^{^{\prime }}a_o(k,t_1)+L_k^{\prime
}c_o^{\dagger }(k,t_1)+M_k^{\prime }b_{odd}(t_1)
\end{equation}

\noindent where the coefficients $A_k^{\prime },B_k^{\prime },F_k^{\prime
},G_k^{\prime },L_k^{\prime },M_k^{\prime }$ are defined as in (\ref{coef})
replacing {\sl even }by {\sl odd. }For the anticommutation relations 
\[
\{\tilde a_e(k,t),\tilde a_e^{\dagger }(k^{\prime },t)\}=\delta _{kk^{\prime
}} 
\]

\[
\{\tilde c_e(k,t),\tilde c_e^{\dagger }(k^{\prime },t)\}=\delta _{kk^{\prime
}} 
\]

\noindent to be valid we should have 
\[
A_k^{*}A_k+B_k^{*}B_k+F_k^{*}F_k=1 
\]

\[
G_k^{*}G_k+L_k^{*}L_k+M_k^{*}M_k=1 
\]

\noindent (no summation over $k$). Both of them are satisfied due to the
fact that the set of wavefunctions $u_{(+)e}(k,x),u_{(-)e}(k,x),u_{even}(x)$
form an orthonormal set. Similar relations are valid for the primed
coefficients. We now make some rather drastic approximations. Since the
potential is assumed initially to be just subcritical the even state lies
very near the negative energy continuum. Also the change in the potential is
assumed to be small so that we end up with a slightly supercritical
potential. Thus it is natural to assume that there is negligible overlap
between the new positive energy wavefunctions on the one hand and the old
negative energy wavefunctions and the bound state on the other. Hence in (%
\ref{coef}) $A_k=B_k=F_k=0$ and 
\begin{equation}
\tilde a_e(k,t_1)=a_e(k,t_1)  \label{m3}
\end{equation}

\begin{equation}  \label{m4}
\tilde a_e^{\dagger }(k,t_1)=a_e^{\dagger }(k,t_1)
\end{equation}

By the same token we can take $G_k=0.$ Thus 
\begin{equation}
\tilde c_e(k,t_1)=M_k^{*}d_b(t_1)+L_k^{*}c_e(k,t_1)\text{ }  \label{m1}
\end{equation}
\begin{equation}
\tilde c_e^{\dagger }(k,t_1)=M_kd_b^{\dagger }(t_1)+L_kc_e^{\dagger }(k,t_1)
\label{m2}
\end{equation}
Note that in the approximation we are working (neglecting the overlap
integrals of the bound state with new states lying in the positive energy
continuum), i.e. essentially assuming completeness of the set $\tilde
u_{(-)e}(k,x)$, and using the orthogonality property of the above set we get 
\begin{equation}
\sum_kM_kM_k^{*}=1  \label{ortho}
\end{equation}

\noindent Following a similar reasoning we deduce that all odd operators
referring to $t=t_1-$ are equal to the corresponding odd operators referring
to $t=t_1+$.

Having obtained the relations beteen operators we come to the definition of
the state $\left| S\right\rangle $ of the system in terms of operators
pertaining to $t=t_1-.$ Recall that since we work in the Heisenberg picture
the state vector is time independent however the vacuum (as all basis kets)
is time dependent. The statement that there is one positron occupying the
even state is equivalent to 
\begin{equation}  \label{dsub}
d^{\dagger }(t_1)d(t_1)\left| S\right\rangle =\left| S\right\rangle
\end{equation}

\noindent We also assume that there are no free electrons or positrons, i.e. 
\begin{equation}
a_{e,o}(k,t_1)\left| S\right\rangle =c_{e,o}(k,t_1)\left| S\right\rangle =0
\label{s6}
\end{equation}

\noindent As already mentioned we can arrange the parameters so that when
the first even state merges with the negative continuum only one (odd) bound
state exists. Although it does not matter in what follows and simply in
order to fix the notation we have to decide whether this odd state is above
or below zero when the potential assumes its critical value. We take the
first possibility; this is indeed the case for the value $a=.7$ mentioned in
the previous section. We assume that there is an electron vacancy in the odd
state, i.e. 
\begin{equation}
b_{odd}(t_1)\left| S\right\rangle =0  \label{ss6}
\end{equation}

Relations (\ref{dsub}), (\ref{s}), (\ref{ss}) specify the state completely.
Using the above equations together with (\ref{m1}), (\ref{m2}) we can
calculate the average number of positrons with momentum $k$ for the
supercritical potential: 
\begin{equation}  \label{num}
N_{e,k}=\left\langle S|\tilde c_e^{\dagger }(k)\tilde c_e(k)|S\right\rangle
=M_kM_k^{*}
\end{equation}

\noindent (\ref{ortho}) gives 
\begin{equation}
\sum_kN_{e,k}=1  \label{numtot}
\end{equation}
i.e. there is one positron emitted. Expression (\ref{num}) for $N_{e,k}$
yields information on the energy spectrum of the positron emitted. Notice
that the mixing between creation and annihilation operators at $t=t_1$, the
ordering of the operators in (\ref{num}) and the use of (\ref{ortho}) are
crucial.

It should be clear from the above discussion that if the original bound
state is filled by an electron then 
\[
d(t_1)\left| S\right\rangle =0 
\]
and the above reasoning leads to the conclusion that no positron is emitted.

We should check whether positron emission as described above is consistent
with charge conservation. The Hamiltonian appropriate for the time dependent
well considered here is given by (\ref{ham}) hence the total charge defined
by (\ref{ch1}) is indeed conserved. To demonstrate this we have to evaluate
the charge expectation value $\left\langle S\right| Q$ $\left|
S\right\rangle $ with $Q$ expressed in the form (\ref{ch}) in terms of the ($%
\sim )$ operators pertaining to $t_1+$ and compare it to $\left\langle
S\right| Q$ $\left| S\right\rangle $ evaluated at $t=t_1-$. At $t_1+$

\begin{equation}  \label{qtotm}
\begin{array}{c}
Q=\sum_k\{-\tilde a_e^{\dagger }(k,t_1)\tilde a_e(k,t_1)+\tilde c_e^{\dagger
}(k,t_1)\tilde c_e(k,t_1)-\tilde a_o^{\dagger }(k,t_1)\tilde a_o(k,t_1)+ \\ 
\\ 
+\tilde c_o^{\dagger }(k,t_1)\tilde c_o(k,t_1)-\tilde b_{odd}^{\dagger
}(t_1)\tilde b_{odd}(t_1)\}+Q_{0,\sup er}
\end{array}
\end{equation}

\noindent and at $t_1-$%
\begin{equation}
\begin{array}{c}
Q=\sum_k\{-a_e^{\dagger }(k,t_1)a_e(k,t_1)+c_e^{\dagger
}(k,t_1)c_e(k,t_1)-a_o^{\dagger }(k,t_1)a_o(k,t_1)+ \\ 
\\ 
+c_o^{\dagger }(k,t_1)c_o(k,t_1)-b_{odd}^{\dagger
}(t_1)b_{odd}(t_1)+d_{even}^{\dagger }(t_1)d_{even}(t_1)\}+Q_{0,sub}
\end{array}
\label{qtotp}
\end{equation}
Using (\ref{numtot}) we notice that the increase of the $\tilde c^{\dagger
}\tilde c$ contribution in the evaluation of the charge expectation value by
one in (\ref{qtotm}) exactly balances the contribution of the $d^{\dagger }d$
in (\ref{qtotp}), thus the contribution of the normal ordered charge $%
Q_{norm}$ rests unchanged. We also observe that $Q_{0,sub}=Q_{0,\sup er}$
when the potential becomes supercritical: An energy level crosses $-m$ but
this does {\sl not} change the spectral asymmetry. (The latter would change
only in the case where the energy level crosses {\sl zero} but this does not
happen at $t=t_1.$) One can see this either from (\ref{qo}) by inspection or
in more detail from (\ref{qoph}): when a bound state merges with the
negative energy continuum $N_{-}$ decreases by 1 but $\delta (-m)$ increases
by $\pi $, the net result being zero.

Regardless of the experimental feasibility we can in principle arrange for
the potential to achieve the supercritical value {\sl and then remain fixed}%
. We would then observe one positron emitted while $V$ goes from $V_{sub}$
to $V_{\sup er}$ and {\sl nothing else}. Thus the transition from a
subcritical to a supercritical potential creates just one positron in this
case and not an electron-positron pair as Zeldovich and Popov (and others)
predict. We shall see presently how and in what sense an additional electron
appears.

At time $t_1+$ the positron charge density is concentrated inside the well
and subsequently leaks away. Since the ex bound state has been transformed
to a transmission resonance the characteristic time scale for the process is
related to the corresponding time delay. Since we intend to switch back to a
subcritical potential at time $t_2$ the time interval $t_2-t_1$ should be
rather large compared to the time delay so that the positron has time to
escape. An estimate of the lifetime is provided in Appendix A.

When we switch back to the subcritical value the positron has escaped from
the well and its charge is carried by travelling waves, hence the
expectation value $\left\langle S|c_e^{\dagger }(k)c_e(k)|S\right\rangle $
has increased by 1 compared to its value for $t<t_1$. We have already shown
that $Q_0$ stays unchanged when a level crosses the threshold $-m$. Thus
charge conservation requires that there is {\sl no }positron occupying the 
{\sl even }bound state for $t>t_2.$ When the potential returns to its
original value $V_{sub}$ the absence of a bound positron signifies the
presence of a bound electron. Thus the round trip $V_{sub}$ $\longrightarrow 
$ $V_{\sup er}$ $\longrightarrow $ $V_{sub}$ starting with an empty electron
bound state results to the emission of one positron and the creation of one
bound electron. It is only in that sense that one may talk about pair
creation. It should be clear that the roles of positron and electron are
rather unequal, the former being free and the latter being bound. Their
roles would of course be reversed for a potential of a different sign. It
must be stressed again that if the bound state were filled by a positron
when the potential is slightly subcritical then {\sl nothing }would happen
when the potential becomes supercritical.

If one prepares the next (higher lying) odd bound state to be empty then as
the potential becomes even stronger, this state merges with the negative
energy continuum when 
\[
V=V_2^c=\sqrt{\frac{\pi ^2}{a^2}+m^2}+m 
\]

\noindent and there is one further positron emitted. On the return trip the
odd bound state is occupied by an electron.\newpage\ 

\section{Conclusions}

We hope that we have now shown how to write down a second quantised theory
of spontaneous fermion production in the context of a Dirac particle bound
by an external field. Dirac's own understanding of a physical positron state
as an unfilled electron state of negative energy is robust and perfectly
capable of dealing with this phenomenon. Considerations of the vacuum charge
(and the related topic of Levinson's theorem) are required for a full
description of the problem. Zeldovich and Popov's intuition which allowed
them to give a detailed analysis of spontaneous positron production based
solely on the first quantised Dirac equation is generally sound. The one
place it fails is in its identification of the cause of positron production
with electron-positron pair production. This identification can be made when
the supercritical potential is switched on at early times and off at late
times, but when the potential is switched on and remains on then spontaneous
positron production is distinct from electron-positron pair production.

We have not shown why narrow positron peaks are observed experimentally in
heavy ion collisions [17]; in principle the overlap integral (6.20) taken
between realistic Coulomb wave functions should provide a clue. A detailed
knowledge of the time-dependence of the external potential is, however,
probably more important. In this paper, we only assume a trivial time
dependence in order to demonstrate the fundamental principles of spontaneous
positron radiation, but we hope to return to the more general problem of
treating electron-positron pair production in a time-dependent external
field in a forthcoming paper.

\newpage\ 

\appendix

\section{Appendix A.}

According to nonrelativistic quantum mechanics the time delay $\Delta t$ of
a wavepacket built around a transmission resonance is given by 
\[
\Delta t=\frac 1{v_0}\frac{d\delta }{dk} 
\]

\noindent where $v_0$ is the group velocity of the wavepacket, $k$ the
wavevector corresponding to the resonance energy and $\delta $ the phase
shift (see e.g. \cite{bohm}). In the present case $v_0$ is the relativistic
group velocity 
\[
v_0=\frac k{\left| E\right| }\text{ , }E^2-k^2=m^2. 
\]
The phase shift is given by expression (\ref{delta}) 
\begin{equation}
\delta (E)=\arctan \left( \frac{1+\gamma ^2}{2\gamma }\tan 2pa\right) -2ka
\label{apdel}
\end{equation}

\noindent where 
\[
\gamma =\frac kp\frac{E+V+m}{E+m} 
\]

\noindent and $p=\sqrt{(E+V)^2-m^2}$ . We are interested in the $N=1$
transmission resonance, hence $p$ satisfies 
\begin{equation}
2pa=\pi  \label{apres}
\end{equation}

From (\ref{apdel}) we get 
\begin{equation}  \label{apdd}
\frac{d\delta }{dk}=a\left( \gamma +\frac 1\gamma \right) \frac{dp}{dk}-2a
\end{equation}

\noindent where we have used (\ref{apres}). Using the expressions for $p$
and $k$ in terms of $E$ we readily calculate the derivative 
\[
\frac{dp}{dk}=\frac kp\frac{E+V}E 
\]

Then after some algebra 
\begin{equation}  \label{apgg}
\left( \gamma +\frac 1\gamma \right) \frac{dp}{dk}=\frac{E+V}E\left( \frac{%
E-m}{E+V-m}+\frac{E+m}{E+V+m}\right)
\end{equation}

\noindent The above expression is exact. We now make the approximation of a
just supercritical potential. Hence we can replace $E$ by $-m$ and $V$ by $%
V_1^c$ as given by (\ref{v1cr}). Substituting (\ref{apgg}) in (\ref{apdd})
we deduce 
\[
\frac{d\delta }{dk}=2a\dfrac m{\sqrt{\dfrac{\pi ^2}{4a^2}+m^2}-m} 
\]

\end{document}